\documentclass[sigconf,nonacm]{acmart}
\usepackage{pgfplots}
\usepackage{tcolorbox}
\pgfplotsset{compat=1.18}
\usepackage{comment}
\usepackage[utf8]{inputenc}
\usepackage[T1]{fontenc}
\usepackage{longtable}
\usepackage{array}     
\usepackage{booktabs}  
\usepackage{tikz}
\usepackage{enumerate}
\usepackage{enumitem}
\usepackage{float}
\usepackage{colortbl}
\usepackage{xcolor}
\usepackage{subcaption}
\usepackage{placeins}
\usepackage{ragged2e}
\usetikzlibrary{patterns}
\usetikzlibrary{positioning, fit,arrows.meta}
\usetikzlibrary{arrows.meta, positioning}

\renewcommand\footnotetextcopyrightpermission[1]{} 

\setcopyright{none}

\AtBeginDocument{
    
}

\begin{document}

\title{Comparing Large Language Models on Scrum Certification-Style Questions: Accuracy, Stability, and Error Patterns}



\author{%
  \href{mailto:robson.vilar@virtus.ufcg.edu.br}{Robson Alves Vilar}, 
  \href{mailto:emanuel.dantas@virtus.ufcg.edu.br}{Emanuel Dantas Filho}, 
  \href{mailto:ademar.sousa@virtus.ufcg.edu.br}{Ademar França de Sousa Neto}, 
  \href{mailto:mirko@virtus.ufcg.edu.br}{Mirko Perkusich}, 
  \href{mailto:danyllo.albuquerque@virtus.ufcg.edu.br}{Danyllo Wagner Albuquerque}, 
  \href{mailto:joao.paiva@virtus.ufcg.edu.br}{João Paiva}, 
  \href{mailto:kyller@virtus.ufcg.edu.br}{Kyller Gorgônio}, 
  \href{mailto:perkusich@virtus.ufcg.edu.br}{Angelo Perkusich}%
}

\affiliation{%
  \vspace{0.3cm}
  \institution{VIRTUS Research, Development and Innovation Center, Federal University of Campina Grande}
  \city{Campina Grande}
  \state{PB}
  \postcode{58429-900}
  \country{Brazil}
}


\renewcommand{\shortauthors}{}
\renewcommand{\shorttitle}{Comparing LLMs on Scrum Certification-Style Questions}

\FloatBarrier
\begin{abstract}
Large Language Models (LLMs) are increasingly used in exam- and certification-style question answering tasks, where their ability to retrieve, interpret, and apply domain-specific knowledge can be systematically assessed. In Software Engineering, such settings are particularly relevant when questions depend on strict adherence to normative definitions, roles, artifacts, and rules. This paper evaluates the performance of three contemporary LLMs, \textit{GPT-5 mini}, \textit{Gemini 3 Flash}, and \textit{DeepSeek Chat 3.2}, in answering 993 Scrum certification-style questions aligned with the Professional Scrum Master I (PSM I) assessment format. We evaluated the models under three prompting strategies (\textit{zero-shot}, \textit{chain-of-thought}, and \textit{source-grounded}), with repeated executions to assess intra-model stability. We also analyzed performance across Scrum topics and question formats, complemented by a qualitative analysis of recurring error patterns in incorrect answers. Results revealed clear differences among models, with Gemini 3 Flash achieving the highest accuracy, followed by GPT-5 mini and DeepSeek Chat 3.2, while intra-model variability remained low across all conditions. By question format, the models achieved the highest accuracy on single-answer multiple-choice items, whereas multi-select and True/False questions were more error-prone. By topic, performance was more consistent in normatively explicit areas such as Artifacts, Empiricism, and Product Value, but more fragile in Scrum Values, Self-Managing Teams, and Stakeholders \& Customers. The qualitative analysis showed that errors were systematic rather than random, involving overgeneralization, restrictive wording, compound distractors, and conflicts between common market interpretations and strict Scrum definitions.
\end{abstract}


\frenchspacing
\maketitle

\FloatBarrier
\section{Introduction}
\label{sec:intro}
Agile Software Development has become a dominant approach to building and evolving software systems across industries. Its principles of collaboration, adaptation, and continuous improvement have shaped how software teams organize work, deliver value, and respond to change \cite{patrucco2022scrum}. As Agile adoption has grown, so has the demand for education and certification mechanisms that support a shared understanding of Agile values, frameworks, and practices \cite{crisostomo2024enhancing}. Scrum is a prominent example of this movement. Certifications such as the \textit{Professional Scrum Master} (PSM) and \textit{Professional Scrum Product Owner} are widely used to assess practitioners' understanding of the Scrum framework and its underlying principles \cite{montenegro2019competences}.

Scrum is particularly well-suited to certification-style assessment because it is grounded in a concise yet normative body of knowledge. The \emph{Scrum Guide} (2020) specifies accountabilities, events, artifacts, commitments, and rules that practitioners are expected to interpret consistently. As a result, Scrum certification-style questions provide a rigorous and measurable setting for assessing domain-specific knowledge whose correctness depends not only on semantic plausibility but also on strict adherence to normative definitions, role boundaries, and rule-based terminology.

In parallel, Large Language Models (LLMs) have been increasingly investigated as question-answering systems in exam- and certification-style settings, where their ability to retrieve, interpret, and apply domain-specific knowledge can be assessed in a controlled manner \cite{Viegas_Gheyi_Ribeiro_2025,inoshita2024assessing}. Recent studies have examined LLM performance in real assessment contexts, including Brazilian exams such as ENADE, ENEM, and POSCOMP, reporting promising results in question answering and exam-solving tasks \cite{mendonca2024enade,piresetal2023enem,Viegas_Gheyi_Ribeiro_2025}. However, strong performance in broad assessment settings does not necessarily imply reliability in domains governed by strict definitions and prescriptive rules. LLMs may still produce plausible but incorrect answers, and their behavior can be sensitive to prompt formulation, raising concerns for certification-style tasks that require precise adherence to domain-specific terminology and normative constraints \cite{Shin2023PromptOrFineTune, SantanaJr2025WhichPromptingTechnique}.

These findings point to the need for investigating LLMs in Software Engineering assessment settings that combine domain specificity, normative precision, and objective scoring, particularly in rule-based certification-style tasks where correctness depends on distinguishing closely related concepts, accountabilities, and normative constraints. This paper addresses that gap through an empirical study of three LLMs (\textit{GPT-5 mini}, \textit{Gemini 3 Flash}, and \textit{DeepSeek Chat 3.2}) on Scrum certification-style questions aligned with the Professional Scrum Master I (PSM I) assessment format. Using a dataset of 993 English questions, we compare model performance across three prompting strategies (\textit{zero-shot}, \textit{chain-of-thought}, and \textit{source-grounded}), analyze variability across repeated executions, and examine how performance varies by Scrum topic and question format. Beyond aggregate accuracy and single-model comparisons, the study includes a qualitative examination of incorrect answers to identify recurring error patterns and distinguish failures associated with model limitations from those related to ambiguity or underspecification in the questions themselves.

Three research questions structure the investigation: how different LLMs perform in terms of accuracy and stability when answering Scrum certification-style questions; how question topic and format influence model performance; and what recurring error patterns emerge in incorrect answers, distinguishing model-side limitations from question-formulation issues.

The main contributions of this paper are:
\begin{enumerate}
\item \textit{A multi-model empirical comparison of contemporary LLMs} on Scrum certification-style questions, considering accuracy and performance differences across models and prompting strategies.
  \item \textit{An analysis of response stability and task sensitivity}, examining intra-model variability across repeated executions and performance differences by Scrum topic and question format.
  \item \textit{A qualitative characterization of error patterns} that helps explain when failures are associated with model limitations and when ambiguous or weakly specified questions influence incorrect answers.
  \item \textit{Empirical implications for evaluating LLMs in normative Software Engineering domains}, showing why aggregate accuracy alone is insufficient and should be complemented by analyses of stability, task characteristics, and recurring failure modes.
\end{enumerate}

The remainder of this paper is organized as follows. Section~\ref{sec:background} presents the background on Scrum as a normative Software Engineering domain, LLMs in Agile and Scrum contexts, and LLMs in exam- and certification-style question answering. Section~\ref{sec:method} presents the method and study protocol. Section~\ref{sec:results} reports the quantitative and qualitative results and discusses their implications for evaluating LLMs in Scrum certification-style tasks. Section~\ref{sec:threats} outlines threats to validity, and Section~\ref{sec:final} concludes with final remarks and future work.

\FloatBarrier
\section{Background}
\label{sec:background}

This section presents the conceptual and empirical background underlying our study. We first discuss Scrum and its certification ecosystem, then review the use of LLMs in Agile and Scrum contexts, and finally summarize evidence on LLMs in exam- and certification-style assessment settings. Together, these areas establish the empirical and conceptual basis for investigating how reliably LLMs answer Scrum certification-style questions.

\textbf{Scrum and Agile Certification}. Scrum is one of the most widely adopted Agile frameworks, emphasizing iterative delivery, self-managing teams, inspection, adaptation, and continuous improvement~\cite{patrucco2022scrum}. Its values, rules, and certification ecosystem are associated with official reference materials such as the \emph{Scrum Guide}, as well as professional certifications including PSM and the Professional Scrum Product Owner (PSPO)~\cite{scrumorg2020}. Because Scrum has a prescriptive and normative structure, it provides a shared vocabulary and explicit rules that support consistent adoption of Agile practices across teams and organizations \cite{patrucco2022scrum}. Scrum certification exams typically rely on multiple-choice questions derived from official materials, offering a controlled setting for assessing knowledge accuracy. The PSM I assessment, in particular, is an introductory certification composed of approximately 80 text-based questions to be completed in 60 minutes, with a minimum score of 85\% required to pass. These characteristics make Scrum certification-style questions an appropriate empirical context for examining how automated systems, including LLMs, interpret and reproduce canonical Agile knowledge \cite{soroka2021importance}.

\textbf{LLMs in Agile and Scrum Contexts}. Recent studies have investigated how LLMs can support Agile software development by improving communication, documentation, automation, and decision-making. Zhang et al.~\cite{zhang2024llm} proposed an autonomous LLM-based agent system to improve user-story quality in industrial Agile teams, reporting gains in consistency and specification accuracy. Cabrero et al.~\cite{cabrero2024exploring} conducted an action research study on the integration of AI assistants into Agile events, showing that such tools can strengthen collaboration and decision-making when appropriate organizational conditions are in place. Other studies have explored cognitive and multi-agent approaches that embed LLMs into Agile workflows. Chudziak et al.~\cite{chudziak2024towards} proposed the CogniSim framework, in which LLM-powered agents assume roles such as developer or reviewer to coordinate Agile tasks. Cinkusz et al.~\cite{cinkusz2024cognitive} extended this perspective to the Scaled Agile Framework, reporting improvements in coordination and communication efficiency. More broadly, Modak et al.~\cite{modak2023integrating} showed that LLMs can accelerate coding, debugging, and documentation; Dhruva et al.~\cite{dhruva2024agile} proposed an LLM-based project management model to improve visibility and delivery velocity; and Samimi et al.~\cite{samimi2025bridging} combined Model-Driven Engineering and Scrum to automate modeling tasks and increase sprint adaptability. Collectively, these studies indicate growing interest in integrating LLMs into Agile practice. However, most of this work emphasizes productivity, collaboration, and process support, leaving less explored whether LLMs can accurately and consistently reproduce the normative, rule-based knowledge codified in the \emph{Scrum Guide}.

\textbf{LLMs in Educational and Certification Assessments}. A third relevant body of work examines LLM performance in formal educational, professional, and certification-style assessments beyond software development and Agile practice. Studies in medicine, business, academic admissions, and computing indicate that modern LLMs can achieve strong results in well-structured, text-based exams. However, their performance is less reliable in the presence of ambiguity, procedural reasoning demands, or visual prompts. For instance, Gerard et al.~\cite{gerard2025evaluating} reported medical-exam performance at or above student level for leading models.
In contrast, Ashrafimoghari et al.~\cite{ashrafimoghari2024evaluating} found that LLMs performed above typical business-school candidates on the GMAT. In computing, Viegas et al.~\cite{Viegas_Gheyi_Ribeiro_2025} showed that several LLMs outperformed human candidates on text-based POSCOMP questions, while still struggling with visual items. Inoshita et al.~\cite{inoshita2024assessing} further demonstrated that prompt techniques can significantly improve accuracy in real estate certification tasks. These findings suggest that LLMs have substantial potential in assessment-like settings, but also reinforce the need to evaluate them in controlled, domain-specific contexts where correctness depends on precise interpretation of rules and terminology.

Despite this progress, little is known about how contemporary LLMs behave in Agile certification contexts, especially in Scrum question answering. Prior work on LLMs in Agile settings has mainly focused on productivity, collaboration, artifact improvement, and workflow automation, rather than on the models' ability to reproduce canonical Scrum knowledge under exam-like conditions. Similarly, studies on LLMs in educational and professional assessments rarely address Software Engineering domains governed by intentionally prescriptive rules and terminology. This is particularly important because, in Scrum certification-style questions, a plausible answer may still be incorrect if it conflicts with the normative definitions of the \emph{Scrum Guide}, ignores restrictive wording, or overgeneralizes a specific accountability.

Building on this context, our study evaluates three contemporary LLMs on 993 Scrum certification-style questions aligned with the PSM I assessment format. Rather than focusing only on aggregate accuracy or on a single model configuration, we analyze performance across models, prompting strategies, repeated executions, Scrum topics, and question formats, and we complement the quantitative results with a qualitative analysis of recurring error patterns. This allows us to provide a more fine-grained view of LLM reliability in a normative Software Engineering domain.


\FloatBarrier
\section{Method}
\label{sec:method}

This section describes the methodological design used to evaluate LLM behavior in Scrum-style question-answering. The study evaluates not only whether LLMs select correct answers, but also whether their behavior remains stable across repeated executions, how performance varies across question characteristics, and which recurring error patterns emerge when their answers diverge from the gold standard. To this end, the method combines controlled model execution, multiple prompting strategies, repeated runs, topic and format-based analyses, and a qualitative examination of incorrect answers.

\subsection{Research Questions}
\label{subsec:rq}

The study is guided by three research questions that progressively address overall model performance, task sensitivity, and qualitative error patterns.

\begin{tcolorbox}[
  colback=black!2,
  colframe=black!35,
  title=Research Questions,
  coltitle=black,
  fonttitle=\bfseries,
  boxrule=0.4pt,
  arc=1.5pt,
  left=4pt,
  right=4pt,
  top=4pt,
  bottom=4pt
]

\textbf{RQ1:} How do different LLMs perform when answering Scrum certification-style questions?\\
\textbf{RQ1.1:} What performance differences can be observed among models?\\
\textbf{RQ1.2:} Is there intra-model variability across repeated executions? \\
\textbf{RQ2:} How do question topic and question format influence LLM performance?\\
\textbf{RQ2.1:} What performance differences can be observed across Scrum topics?\\
\textbf{RQ2.2:} What performance differences can be observed across question formats? \\
\textbf{RQ3:} What error patterns can be observed in LLM answers to Scrum certification-style questions?\\
\textbf{RQ3.1:} Which errors are associated with ambiguity or underspecification in the questions?\\
\textbf{RQ3.2:} Which errors are more strongly associated with intrinsic model limitations?
\end{tcolorbox}

The research questions move beyond aggregate accuracy by examining model performance, response stability, task sensitivity, and recurring error patterns. RQ1 examines overall model behavior, considering both differences in accuracy across models and response stability across repeated executions. RQ2 shifts the analysis to task characteristics, investigating whether performance varies across Scrum topics and question formats. RQ3 complements the quantitative analysis by examining recurring error patterns and distinguishing failures associated with ambiguity or underspecification in the questions from those more closely related to model limitations.

\subsection{Study Design}
\label{subsec:study-design}

\begin{figure*}[t]
\centering
\footnotesize
\includegraphics[width=0.75\textwidth]{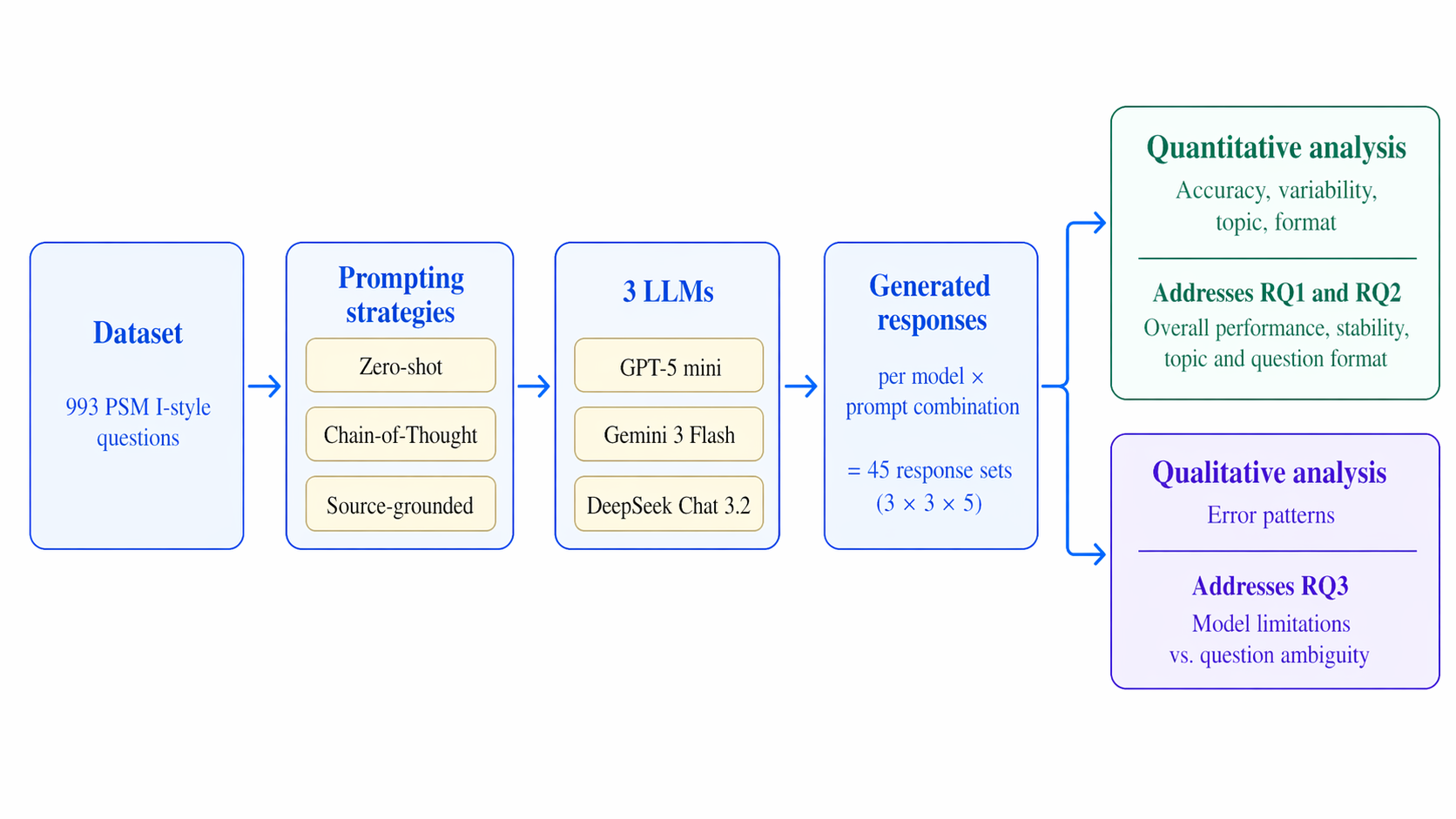}
\caption{Overview of the experimental design.}
\label{fig:design}
\end{figure*}

We conducted a mixed-method empirical study combining controlled LLM execution, quantitative question-response evaluation, and qualitative error analysis, following the guidelines for empirical Software Engineering research proposed by Wohlin et al.~\cite{wohlin2012experimentation} and Ralph et al.~\cite{ralph2020empirical}. Figure~\ref{fig:design} summarizes the experimental pipeline, which starts from a dataset of Scrum certification-style questions, applies different prompting strategies to multiple LLMs, collects generated responses, and then analyzes the results quantitatively for RQ1 and RQ2 and qualitatively for RQ3.

\subsubsection{Dataset}

The dataset consists of 993 English Scrum certification-style questions obtained from a PSM I preparation question bank maintained by an experienced Scrum training provider. The questions have been used and refined across multiple course editions and were designed to reflect the format and content emphasis of the Professional Scrum Master I (PSM I) assessment. They cover the main Scrum knowledge areas defined in the \emph{Scrum Guide} (2020). Each question includes a textual stem, a set of alternatives, a Scrum topic category, and a validated answer key.

The dataset comprises three formats commonly used in certification-style assessments: True/False questions (249 items), single-answer multiple-choice questions (591 items), and multi-select questions with multiple correct options (153 items). All questions were anonymized and stored using a minimal schema (\texttt{question}, \texttt{category}, \texttt{options}, \texttt{gold\_set}). The anonymized dataset schema and the experimental outputs are available as supplementary material. The release of the full question text follows the usage permissions associated with the original question bank.

The full dataset was used for the overall model comparison and repeated-execution analysis associated with RQ1.1 and RQ1.2. For the finer-grained analyses in RQ2, we adopted task-specific balanced subsamples to reduce category imbalance and enable fairer comparisons. Throughout the analyses, the three prompting strategies are abbreviated as ZS (zero-shot), CoT (chain-of-thought), and SG (source-grounded). For topic-based analysis, we used a balanced sample of 104 questions, with eight questions for each of 13 Scrum subject areas: Artifacts, Coaching and Mentoring, Done, Empiricism, Events, Facilitation, Forecasting and Release Planning, Product Backlog Management, Product Value, Scrum Team, Scrum Values, Self-Managing Teams, and Stakeholders \& Customers. For question-format analysis, we used a balanced sample of 96 questions, with 32 questions per format category.

\subsubsection{Prompting Strategies}

We evaluated three prompting strategies that reflect different levels of instruction and source grounding:

\begin{itemize}
  \item \textit{Zero-shot}: the question and its alternatives are presented without additional guidance, serving as a baseline for direct answer generation.
  \item \textit{Chain-of-Thought}: the model is instructed to reason briefly before stating the final answer, encouraging explicit intermediate justification \cite{wei2022chain}.
  \item \textit{Source-grounded}: the model is required to justify its answer by explicitly grounding it in the \emph{Scrum Guide} (2020), promoting alignment with canonical material \cite{lewis2020retrieval}.
\end{itemize}

These strategies were selected to represent common ways of querying LLMs in question-answering scenarios, ranging from direct answering to more structured and source-aware prompts. The same prompting structure was used for all models within each prompting condition, allowing differences in performance to be interpreted primarily in terms of model behavior rather than prompt variation.

\subsubsection{LLMs and Generated Responses}

Each question was answered independently using three LLMs: \textit{GPT-5 mini}, \textit{Gemini 3 Flash}, and \textit{DeepSeek Chat 3.2}, all accessed through their respective APIs. To reduce cross-question contamination, each question was executed in isolation, with no conversational memory between runs. Default model settings were maintained across executions.

To investigate response stability, each model--prompt combination was executed five times over the full dataset. This repeated-execution design enabled the analysis of intra-model variability and reduced the risk of drawing conclusions from a single stochastic run. The generated responses were then validated for schema conformity and normalized to a canonical representation before scoring.

\subsubsection{Quantitative Analyses}

For RQ1, we compared model performance using exact-match accuracy, computed by comparing each predicted answer against the gold-standard answer. Exact-match scoring was also applied to multi-select questions, meaning that a response was considered correct only when the predicted set matched the complete gold-standard set. Average accuracy was computed across the five repeated executions for each model--prompt combination.

To assess RQ1.2, we analyzed intra-model variability across repeated executions using descriptive measures, including standard deviation, minimum and maximum observed accuracy, and run-to-run amplitude. This analysis allowed us to examine whether model performance remained stable across repeated runs under the same experimental conditions.

For RQ2, we analyzed performance by Scrum topic and question format using the balanced subsamples described above. The topic-level analysis examined how accuracy varied across the 13 Scrum subject areas, while the format-level analysis compared True/False, single-answer multiple-choice, and multi-select questions. These analyses were descriptive and comparative, aiming to identify systematic strengths and weaknesses associated with task characteristics.

\subsubsection{Qualitative Analyses}

For RQ3, we complemented the quantitative results with a qualitative analysis of selected incorrect answers, following qualitative synthesis practices commonly used in empirical Software Engineering \cite{cruzes2011recommended}. We first inspected disagreement patterns across the full dataset, focusing on cases in which all models failed under the same prompting condition, as well as cases in which only one model produced the correct answer. Based on this inspection, we selected representative cases for qualitative analysis.

The qualitative analysis considered two broad categories of error: (\textit{i}) errors more strongly associated with intrinsic model limitations, such as overgeneralization, conceptual confusion, difficulty handling restrictive wording, or difficulty managing multiple correct answers; and (\textit{ii}) errors associated with ambiguity, underspecification, or multiple plausible interpretations in the question itself.

The analysis was conducted collaboratively by the two main authors, who reviewed and discussed the selected cases to interpret the likely sources of error. When discrepancies suggested possible issues with the gold standard, the interpretations were further checked with the dataset provider, who served as an additional source to clarify item intent and answer-key consistency. This procedure allowed us to examine not only whether models answered incorrectly, but also what kinds of failure patterns emerged in a normative Software Engineering question-answering task.

\FloatBarrier
\section{Results and Discussion}
\label{sec:results}

This section presents the results of our empirical study of LLM behavior in Scrum-style question-answering. We organize the discussion around three dimensions aligned with the research questions: overall model performance and response stability (RQ1), the influence of Scrum topic and question format (RQ2), and qualitative error patterns in incorrect answers (RQ3). We also discuss what these findings indicate about the evaluation of LLMs in normative Software Engineering knowledge tasks.

\subsection{Accuracy Across Models and Prompts (RQ1.1)}
\label{subsec:rq1}

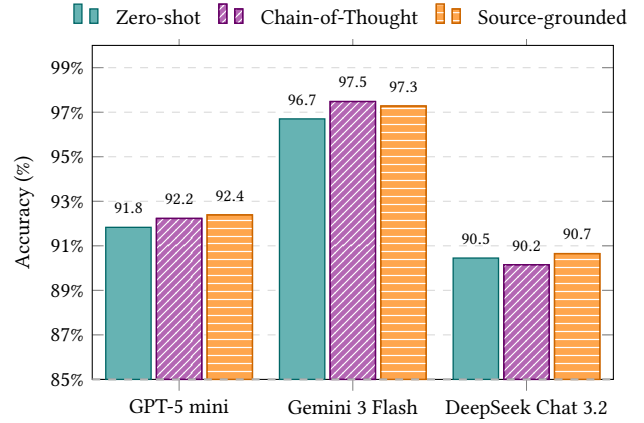
\begin{figure}[t]
\centering
\begin{tikzpicture}
\begin{axis}[
  ybar,
  bar width=0.6cm, 
  width=\columnwidth,
  height=6cm,
  ymin=85, ymax=100,
  ymajorgrids=true,
  grid style={dashed, gray!30},
  xtick=data,
  symbolic x coords={GPT-5 mini, Gemini 3 Flash, DeepSeek Chat 3.2},
  x tick label style={font=\small},
  y tick label style={font=\small},
  ylabel={Accuracy (\%)},
  ylabel style={font=\small},
  ytick={85,87,89,91,93,95,97,99},
  yticklabel={\pgfmathprintnumber{\tick}\%},
  enlarge x limits=0.25,
  legend style={
    at={(0.5,1.02)},
    anchor=south,
    legend columns=3,
    font=\small,
    draw=none,
    column sep=0.5em,
  },
  legend cell align=left,
  clip=false,
  nodes near coords,
  nodes near coords style={
    font=\scriptsize,
    yshift=2pt,
    /pgf/number format/.cd,
      fixed,
      precision=1
  },
  point meta=y
]

\addplot[
  fill=teal!60, 
  draw=teal!80!black, 
  line width=0.5pt
]
coordinates {
  (GPT-5 mini,91.83) 
  (Gemini 3 Flash,96.70) 
  (DeepSeek Chat 3.2, 90.45)
};

\addplot[
  fill=violet!60, 
  draw=violet!80!black, 
  line width=0.5pt,
  postaction={pattern=north east lines, pattern color=white}
]
coordinates {
  (GPT-5 mini,92.23) 
  (Gemini 3 Flash,97.48) 
  (DeepSeek Chat 3.2, 90.15)
};

\addplot[
  fill=orange!70, 
  draw=orange!80!black, 
  line width=0.5pt,
  postaction={pattern=horizontal lines, pattern color=white}
]
coordinates {
  (GPT-5 mini,92.39) 
  (Gemini 3 Flash,97.28) 
  (DeepSeek Chat 3.2, 90.65)
};

\legend{Zero-shot, Chain-of-Thought, Source-grounded}

\draw[dashed, gray!60, line width=0.8pt]
  ({rel axis cs:0,0}|-{axis cs:GPT-5 mini,85}) --
  ({rel axis cs:1,0}|-{axis cs:GPT-5 mini,85});

\end{axis}
\end{tikzpicture}
\caption{Average accuracy (\%) by model and prompting strategy across 993 questions.
The dashed line indicates the PSM~I passing threshold (85\%).}
\label{fig:accuracy_by_model}
\end{figure}

Figure~\ref{fig:accuracy_by_model} presents the average accuracy of the three evaluated LLMs under the three prompting strategies. Overall performance was high across the 993 questions in the dataset, with all models exceeding the PSM~I passing threshold. At the same time, differences among models were clear and consistent across prompting conditions.

Gemini 3 Flash achieved the highest accuracy across all prompting strategies, followed by GPT-5 mini, while DeepSeek Chat 3.2 consistently obtained the lowest scores. These differences were clear across the mean results of the five repeated executions. In practical terms, this indicates that model choice had a stronger effect on answer accuracy than prompt choice in this Scrum certification-style question-answering task.

Prompting produced smaller effects than model selection. GPT-5 mini improved slightly from zero-shot to the more structured prompts, reaching its best result under the source-grounded condition (92.39\%). Gemini 3 Flash also benefited from structured prompting, peaking under Chain-of-Thought (97.48\%), while remaining very close under source-grounded prompting (97.28\%). DeepSeek Chat 3.2 showed only modest variation, with its best result under source-grounded prompting (90.65\%). Overall, these results suggest that prompt engineering can generate incremental improvements, but does not eliminate underlying differences in model capability.

These differences are relevant because high aggregate performance can obscure substantial variation among models. Although all three models exceeded the 85\% PSM~I passing threshold, the gap between approximately 97\% and 90\% accuracy remains practically meaningful. Across a dataset of nearly one thousand questions, this difference corresponds to dozens of additional incorrect answers, reinforcing the need to analyze model-specific behavior rather than treating contemporary LLMs as interchangeable systems.

Structured prompting produced modest gains in some conditions, especially when prompts encouraged reasoning or grounding in the \emph{Scrum Guide}. However, these gains did not substantially change the overall ranking of models, suggesting that prompt design improved performance incrementally but did not compensate for differences in model capability.

\subsection{Intra-model Variability Across Repeated Executions (RQ1.2)}
\label{subsec:rq1_2}

To assess response stability, each model--prompt combination was executed five times over the full dataset. Overall, intra-model variability was low across all evaluated conditions, with standard deviations ranging from 0.100 to 0.583 percentage points and run-to-run ranges from 0.200 to 1.600 percentage points.

Gemini 3 Flash showed the lowest variability overall, particularly in the zero-shot condition (standard deviation = 0.100; range = 0.200), indicating highly stable behavior across repeated runs. DeepSeek Chat 3.2 also remained stable across conditions, with a standard deviation ranging from 0.167 to 0.342 and a range from 0.400 to 0.800. GPT-5 mini showed slightly higher variation, especially under source-grounded prompting (standard deviation = 0.583; range = 1.600), although even this fluctuation remained small in absolute terms.

These results indicate that repeated executions did not substantially alter the study's overall conclusions. The ranking of models remained stable across runs, and no condition exhibited large oscillations that would call the main findings into question. This strengthens the interpretation that the observed differences among models and prompting strategies are not artifacts of isolated stochastic executions.

At the same time, low variability should not be confused with correctness. A model may behave consistently while still being systematically wrong in specific cases, such as misinterpreting restrictive wording, failing on particular question formats, or converging on plausible but incorrect answers in ambiguous items. Thus, response stability should be understood as one dimension of model behavior, not as evidence of uniform accuracy. In this study, the low intra-model variability strengthens the credibility of the subsequent analyses: the performance differences observed across Scrum topics (RQ2.1), question formats (RQ2.2), and recurring error patterns (RQ3) are unlikely to result from isolated run-level fluctuations. This means that the weaknesses identified in later analyses (such as fragility in \textit{Stakeholders \& Customers} or sensitivity to restrictive wording) reflect stable behavioral tendencies rather than stochastic artifacts.

\subsection{Performance by Scrum Topic (RQ2.1)}
\label{subsec:rq2_1}

Table~\ref{tab:full_prompt_results} presents the topic-level accuracy results across the three evaluated models and prompting strategies. This analysis used a balanced topic sample of 104 questions (eight per Scrum subject area), allowing fair comparisons across topics.

\begin{table*}[t]
\centering
\caption{Mean accuracy (\%) for topics with variable or low performance. Topics reaching perfect or near-perfect accuracy across all models and conditions are omitted. Shaded cells indicate values below 80\% (darker = lower).}
\label{tab:full_prompt_results}
\small
\begin{tabular}{lccccccccc}
\toprule
\textbf{Topic} & \multicolumn{3}{c}{\textbf{Gemini 3 Flash}} & \multicolumn{3}{c}{\textbf{GPT-5 mini}} & \multicolumn{3}{c}{\textbf{DeepSeek Chat 3.2}} \\
\cmidrule(lr){2-4} \cmidrule(lr){5-7} \cmidrule(lr){8-10}
 & ZS & CoT & SG & ZS & CoT & SG & ZS & CoT & SG \\
\midrule
Done & 100 & 100 & 100 & 88 & 88 & 88 & 88 & 88 & \cellcolor{gray!25}75 \\
Events & 95 & 100 & 100 & 100 & 100 & 100 & 100 & 100 & 100 \\
Product Backlog Mgmt. & 94 & 94 & 100 & 88 & 100 & 88 & 88 & 88 & 88 \\
Scrum Team & 96 & 88 & 88 & 100 & 100 & 100 & 100 & 100 & 100 \\
Scrum Values & 100 & 100 & 100 & \cellcolor{gray!25}75 & \cellcolor{gray!50}38 & \cellcolor{gray!35}62 & 100 & \cellcolor{gray!35}62 & \cellcolor{gray!25}75 \\
Self-Managing Teams & \cellcolor{gray!50}33 & 100 & 100 & 88 & \cellcolor{gray!25}75 & 88 & \cellcolor{gray!25}75 & 88 & 88 \\
Stakeholders \& Customers & \cellcolor{gray!35}50 & \cellcolor{gray!35}50 & \cellcolor{gray!35}50 & \cellcolor{gray!35}62 & \cellcolor{gray!35}62 & \cellcolor{gray!35}62 & \cellcolor{gray!35}62 & \cellcolor{gray!35}50 & \cellcolor{gray!35}50 \\
\bottomrule
\end{tabular}
\end{table*}

The results show that LLM performance varies substantially across Scrum topics. Six topics, \textit{Artifacts}, \textit{Empiricism}, \textit{Product Value}, \textit{Coaching and Mentoring}, \textit{Facilitation}, and \textit{Forecasting and Release Planning}, reached perfect or near-perfect accuracy across all models and prompting strategies, indicating that the evaluated models handled questions consistently when answers depended on direct recognition of canonical Scrum concepts. Table~\ref{tab:full_prompt_results} focuses on the remaining topics, where performance was lower or more variable.

In contrast, lower and more variable performance was observed in topics that require finer distinctions among responsibilities, values, or stakeholder relationships. The clearest case is \textit{Stakeholders \& Customers}, which showed low accuracy across all models and prompting strategies. \textit{Self-Managing Teams} also showed high sensitivity to prompting in Gemini 3 Flash, increasing from 33\% under zero-shot prompting to 100\% under both Chain-of-Thought and source-grounded prompting. \textit{Scrum Values} exhibited another form of topic-level fragility: while Gemini 3 Flash reached 100\% across all prompts, GPT-5 mini and DeepSeek Chat 3.2 showed considerably lower results in some configurations.

Prompting effects were uneven across topics. In some cases, structured prompts improved performance, as observed for \textit{Self-Managing Teams} with Gemini 3 Flash and for \textit{Product Backlog Management} with source-grounded prompting. However, these improvements were not consistent across all models or topics. For example, GPT-5 mini performed worse on \textit{Scrum Values} under Chain-of-Thought prompting than under zero-shot prompting, and DeepSeek Chat 3.2 also showed decreases in some topic--prompt combinations. Thus, prompting should not be interpreted as uniformly beneficial; its effect appears to depend on the interaction between model, topic, and question wording.

Overall, the topic-level analysis confirms that high aggregate accuracy can mask localized weaknesses. Even when models perform well on average, they may still be fragile in Scrum areas that require distinguishing between similar accountabilities, interpreting value-oriented concepts, or resolving context-dependent stakeholder relationships. These findings motivate the subsequent analysis by question format and the qualitative examination of recurring error patterns.

\subsection{Performance by Question Format (RQ2.2)}
\label{subsec:rq2_2}

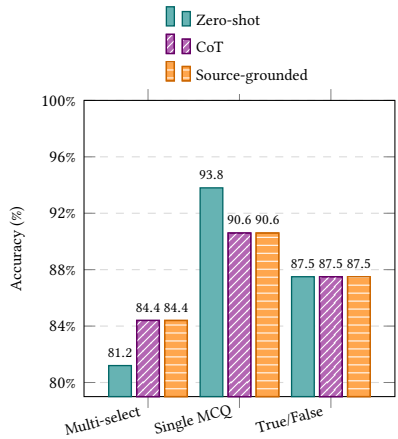
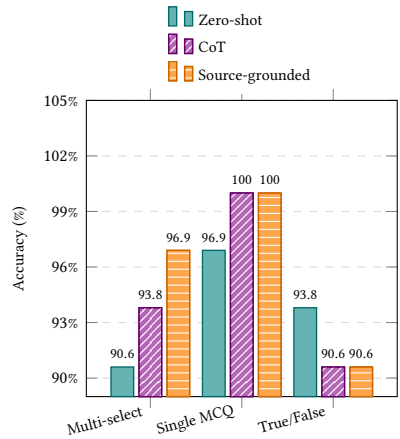
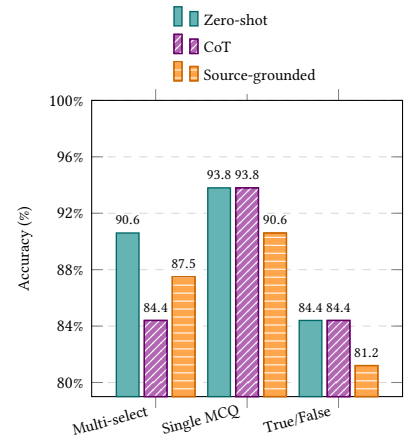
\begin{figure*}[t]
\centering
\begin{subfigure}[t]{0.32\textwidth}
\centering
\begin{tikzpicture}
\begin{axis}[
  ybar=2pt,
  bar width=0.30cm,
  height=5.5cm,
  width=\textwidth,
  ymajorgrids=true,
  grid style={dashed, gray!30},
  xtick=data,
  symbolic x coords={Multi-select, Single MCQ, True/False},
  x tick label style={font=\scriptsize, rotate=15, anchor=east},
  y tick label style={font=\scriptsize},
  ylabel={Accuracy (\%)},
  ylabel style={font=\scriptsize},
  ymin=79, ymax=100,
  ytick={80,84,88,92,96,100},
  yticklabel={\pgfmathprintnumber{\tick}\%},
  enlarge x limits=0.35,
  legend style={at={(0.5,1.02)}, anchor=south, legend columns=1, font=\scriptsize, draw=none},
  legend cell align=left,
  clip=false,
  nodes near coords,
  nodes near coords style={font=\tiny, /pgf/number format/fixed, /pgf/number format/precision=1},
  point meta=y,
]
\addplot[fill=teal!60, draw=teal!80!black, line width=0.5pt]
  coordinates {(Multi-select,81.20) (Single MCQ,93.80) (True/False,87.50)};
\addplot[fill=violet!60, draw=violet!80!black, line width=0.5pt,
         postaction={pattern=north east lines, pattern color=white}]
  coordinates {(Multi-select,84.40) (Single MCQ,90.60) (True/False,87.50)};
\addplot[fill=orange!70, draw=orange!80!black, line width=0.5pt,
         postaction={pattern=horizontal lines, pattern color=white}]
  coordinates {(Multi-select,84.40) (Single MCQ,90.60) (True/False,87.50)};
\legend{Zero-shot, CoT, Source-grounded}
\end{axis}
\end{tikzpicture}
\caption{GPT-5 mini}
\label{fig:fmt_gpt}
\end{subfigure}
\hfill
\begin{subfigure}[t]{0.32\textwidth}
\centering
\begin{tikzpicture}
\begin{axis}[
  ybar=2pt,
  bar width=0.30cm,
  height=5.5cm,
  width=\textwidth,
  ymajorgrids=true,
  grid style={dashed, gray!30},
  xtick=data,
  symbolic x coords={Multi-select, Single MCQ, True/False},
  x tick label style={font=\scriptsize, rotate=15, anchor=east},
  y tick label style={font=\scriptsize},
  ylabel={Accuracy (\%)},
  ylabel style={font=\scriptsize},
  ymin=89, ymax=105,
  ytick={90,93,96,99,102,105},
  yticklabel={\pgfmathprintnumber{\tick}\%},
  enlarge x limits=0.35,
  legend style={at={(0.5,1.02)}, anchor=south, legend columns=1, font=\scriptsize, draw=none},
  legend cell align=left,
  clip=false,
  nodes near coords,
  nodes near coords style={font=\tiny, /pgf/number format/fixed, /pgf/number format/precision=1},
  point meta=y,
]
\addplot[fill=teal!60, draw=teal!80!black, line width=0.5pt]
  coordinates {(Multi-select,90.60) (Single MCQ,96.90) (True/False,93.80)};
\addplot[fill=violet!60, draw=violet!80!black, line width=0.5pt,
         postaction={pattern=north east lines, pattern color=white}]
  coordinates {(Multi-select,93.80) (Single MCQ,100.00) (True/False,90.60)};
\addplot[fill=orange!70, draw=orange!80!black, line width=0.5pt,
         postaction={pattern=horizontal lines, pattern color=white}]
  coordinates {(Multi-select,96.90) (Single MCQ,100.00) (True/False,90.60)};
\legend{Zero-shot, CoT, Source-grounded}
\end{axis}
\end{tikzpicture}
\caption{Gemini 3 Flash}
\label{fig:fmt_gemini}
\end{subfigure}
\hfill
\begin{subfigure}[t]{0.32\textwidth}
\centering
\begin{tikzpicture}
\begin{axis}[
  ybar=2pt,
  bar width=0.30cm,
  height=5.5cm,
  width=\textwidth,
  ymajorgrids=true,
  grid style={dashed, gray!30},
  xtick=data,
  symbolic x coords={Multi-select, Single MCQ, True/False},
  x tick label style={font=\scriptsize, rotate=15, anchor=east},
  y tick label style={font=\scriptsize},
  ylabel={Accuracy (\%)},
  ylabel style={font=\scriptsize},
  ymin=79, ymax=100,
  ytick={80,84,88,92,96,100},
  yticklabel={\pgfmathprintnumber{\tick}\%},
  enlarge x limits=0.35,
  legend style={at={(0.5,1.02)}, anchor=south, legend columns=1, font=\scriptsize, draw=none},
  legend cell align=left,
  clip=false,
  nodes near coords,
  nodes near coords style={font=\tiny, /pgf/number format/fixed, /pgf/number format/precision=1},
  point meta=y,
]
\addplot[fill=teal!60, draw=teal!80!black, line width=0.5pt]
  coordinates {(Multi-select,90.60) (Single MCQ,93.80) (True/False,84.40)};
\addplot[fill=violet!60, draw=violet!80!black, line width=0.5pt,
         postaction={pattern=north east lines, pattern color=white}]
  coordinates {(Multi-select,84.40) (Single MCQ,93.80) (True/False,84.40)};
\addplot[fill=orange!70, draw=orange!80!black, line width=0.5pt,
         postaction={pattern=horizontal lines, pattern color=white}]
  coordinates {(Multi-select,87.50) (Single MCQ,90.60) (True/False,81.20)};
\legend{Zero-shot, CoT, Source-grounded}
\end{axis}
\end{tikzpicture}
\caption{DeepSeek Chat 3.2}
\label{fig:fmt_deepseek}
\end{subfigure}
\caption{Accuracy (\%) by question format and prompting strategy across the three evaluated models.}
\label{fig:fmt_all}
\end{figure*}

To examine the effect of question format, we used the balanced sample of 96 questions described in Section~\ref{subsec:study-design}, with 32 questions per format category. Figure~\ref{fig:fmt_all} presents the accuracy results for each model across question format and prompting strategy.

Across models and prompts, single-answer multiple-choice questions achieved the highest accuracy, with an aggregated mean of approximately 94.45\%. Multi-select questions averaged 88.2\%, while True/False questions showed the lowest overall performance, at approximately 87.5\%. This pattern indicates that question format influences model performance even when the topic domain and scoring procedure remain fixed.

Model-specific differences remained visible within this general pattern. GPT-5 mini showed its lowest results on multi-select questions, suggesting sensitivity to formats that require identifying a complete set of correct alternatives. Gemini 3 Flash achieved the strongest results across all formats, reaching 100\% accuracy on single-answer multiple-choice questions under both Chain-of-Thought and source-grounded prompting. DeepSeek Chat 3.2 performed worst on True/False questions, especially under source-grounded prompting.

These findings show that aggregate accuracy can mask format-specific weaknesses. Single-answer multiple-choice questions appear to be the most favorable format for the evaluated models, whereas multi-select and True/False questions introduce additional difficulty. In multi-select items, the model must avoid both missing correct alternatives and selecting plausible distractors; in True/False items, a single restrictive term or subtle normative distinction can invert the expected answer. Question format is, therefore, a meaningful variable in certification-style evaluations of LLMs, one that aggregate scoring alone cannot capture.

\subsection{Errors Stemming from Model Limitations (RQ3.1)}
\label{subsec:rq3_1}

The previous analyses showed that the evaluated LLMs achieved high aggregate accuracy, but also revealed variation across models, topics, and question formats. To better understand these results, this subsection examines recurring error patterns in incorrect answers that are more closely associated with model behavior, specifically, overgeneralization, difficulty handling restrictive wording, and incorrect treatment of compound alternatives. The analysis was conducted collaboratively by the two main authors using traceable question identifiers and explicit analytic categories, with cross-checking by the dataset provider for cases involving possible gold-standard inconsistencies.

The first group of errors does not resemble hallucination in the narrow sense of inventing facts. Instead, the selected cases suggest difficulties in handling restrictive wording, strong semantic associations, and compound alternatives. Three questions illustrate these recurring limitations.

The first error pattern involves semantic association overriding modal qualification, as illustrated by Question 280:

\begin{quote}
\textbf{Question 280 -- True/False}\\
True or False: The Product Backlog might commit to a Product Goal.\\
1) TRUE\\
2) FALSE\\
\textit{Correct answer: 2) FALSE}
\end{quote}

In this scenario, all three models, across all prompt approaches, incorrectly classified the statement as "TRUE". The justification generated by the models, revealed especially in the Chain-of-Thought (CoT) prompts, demonstrates that they successfully retrieved the correlation between the artifacts ("Product Backlog") and their respective commitments ("Product Goal"), a novelty introduced in the 2020 Scrum Guide. However, the error appears to result from insufficient weighting of the modal verb ``might'' when compared with the strong semantic association among the main terms.

In this case, the models appear to prioritize the strong association between \textit{Product Backlog}, \textit{commitment}, and \textit{Product Goal}, while underweighting the restrictive effect of the modal verb ``might''. As a result, they treat a mandatory relationship in Scrum as if it were compatible with mere possibility. The error, therefore, reflects difficulty in handling fine-grained logical qualification, rather than a failure to retrieve the relevant Scrum concepts.

The second pattern involves the influence of common market interpretations over strict normative definitions, as illustrated by Question 675:

\begin{quote}
\textbf{Question 675 -- True/False}\\
True or False: The Product Owner is the main person responsible for engaging the stakeholders\\
1) TRUE\\
2) FALSE\\
\textit{Correct answer: 2) FALSE}
\end{quote}

Again, all three models selected the incorrect answer. The official answer key indicates that the statement is false, based on the 2020 Scrum Guide, which assigns the responsibility for engaging with stakeholders to the entire Scrum Team, removing the Product Owner from the position of an exclusive or "main" intermediary. However, LLMs are trained on vast corpora of data collected from the internet, which encompass over a decade of agile forums, blogs, old preparatory courses, and practical market discussions. In the "common sense" of the technology market, the Product Owner is, in fact, expected and taught to be the main point of contact with the client and stakeholders.

This case suggests difficulty in prioritizing the strict normative definition of Scrum over widespread informal interpretations of the Product Owner role. Even when the Source-grounded technique was applied to direct the models' attention to the Scrum Guide roles, the pre-training data's probabilistic bias persisted. The models used their text-generation capabilities to create fallacious rationalizations (for example, arguing that, since the PO manages the Backlog, they are, by logical consequence, the primary person responsible for the stakeholders). This illustrates a significant limitation of LLMs in certification exams that require textual rigor: their tendency to favor majority knowledge from the public domain over recent, stringent normative updates.

The third pattern concerns difficulty with compound distractors in multiple-choice questions, as observed in Question 556:

\begin{quote}
\textbf{Question 556 -- Multi-select}\\
When should Product Backlog items be refined? (Choose the best two answers.)\\
1. Only in the first Sprint\\
2. During Sprint Planning events\\
3. Throughout the Sprints\\
4. During the Sprint, if they have not been refined in the previous Sprints\\
\textit{Correct answer: 2, 3}
\end{quote}

In this question, the models consistently opted for options 3 and 4, missing the correct answer. Option 4 is a classic complex distractor, built on a true premise (``During the Sprint'') and a false limiting condition (``if they have not been refined in the previous Sprints''). The Scrum Guide is clear in stating that refinement is an ongoing activity that occurs "as needed". This error illustrates the difficulty of evaluating compound propositions in multiple-choice questions. Models tend to approve an alternative if the first clause has high semantic similarity to the reference documents, often failing to invalidate it based on a subsequent restrictive conditional clause. Furthermore, the models rejected option 2 (``During Sprint Planning events'') due to a negative association bias: in the Scrum data ecosystem, there is a massive emphasis on teaching that refinement``is an ongoing activity, not a formal event''. The term ``event'' in option 2 may have discouraged the models from recognizing that refinement can also occur during Sprint Planning when needed.

\subsection{Errors Associated with Ambiguity and Question Formulation (RQ3.2)}
\label{subsec:rq3_2}

The errors examined in RQ3.1 were primarily associated with how models process and weight information, overgeneralizing semantic associations, underweighting modal qualifiers, or failing to evaluate compound alternatives. RQ3.2 shifts the focus to the questions themselves, examining how ambiguity, underspecification, or competing interpretations in the item formulation can also lead to incorrect answers, even when the model's reasoning is locally coherent. In this group of cases, ambiguity, underspecification, or competing interpretations of the reference material make the expected answer less straightforward. The most prominent example involves a tension between a general Scrum rule and a more specific statement associated with the Sprint Retrospective, as illustrated by Question 61:

\begin{quote}
\textbf{Question 61 - Multiple choice}\\
During the Sprint Retrospective, the Scrum Team members identify the most helpful changes to improve their effectiveness. Where are the most impactful improvements registered? (choose the best answer)\\
1) Next Sprint Backlog\\
2) Issue Tracker\\
3) Product Backlog\\
4) Process Improvement Backlog\\
\textit{Correct answer: 3) Product Backlog}
\end{quote}

In this case, all models selected option 1 (Next Sprint Backlog). Rather than indicating a straightforward model limitation, this error appears to reflect ambiguity in how the question relates two Scrum Guide statements. The author of the question based the answer key (Product Backlog) on the framework's macro rule: the Product Backlog section unequivocally states that it is the "sole source of work undertaken by the Scrum Team". The logic follows that any improvement requiring work must reside there. However, in the section dedicated to the Sprint Retrospective, the guide establishes an exception or specific rule: "The most impactful improvements are addressed as soon as possible. They may even be added to the Sprint Backlog for the next Sprint."

The formulation of Question 61 appears to have guided the models toward the more context-specific rule. By including the opening phrase "During the Sprint Retrospective...", the prompt activated the attention weights directed at the scope of that specific ceremony in the model's knowledge vector space. As a result, the LLM searched for the rule most semantically proximal to the Retrospective event, retrieving the explicit instruction that authorizes the registry in the "Next Sprint Backlog". In this sense, the models selected a semantically defensible answer, even though it diverged from the adopted gold standard. The question is inherently ambiguous, as it does not specify which hierarchical level of rule (the general law of the Product Backlog or the Retrospective's exception) should be applied. This case shows that some incorrect answers may be linked to ambiguity or underspecification in the question, rather than to a simple failure to retrieve Scrum knowledge.

A second formulation-related pattern involves restrictive linguistic markers. This connects Questions 280 and 675 in terms of question wording. Terms such as ``might'' and ``main'' can change the truth value of an otherwise plausible statement, requiring precise attention to the wording of the item and to the terminology of the Scrum Guide.

In Question 675, the inclusion of the restrictive word "main" transforms a premise that is almost true in daily practice into a false statement in the exam environment. Creators of certification tests frequently introduce adverbs of intensity or exclusivity (main, solely, only, strictly) to invalidate statements that are instinctively correct for experienced professionals. This tricky formulation style exploits human cognitive heuristics of speed reading. The results suggest that LLMs can be similarly vulnerable to such wording effects. An ambiguous formulation, or one that uses hyper-specific lexical nuances, acts as a vector of confusion, forcing the model to weigh the semantic probability of the entire sentence against a single pivoting word. Most of the time, as tests consistently demonstrated, the model fails to assign sufficient weight to the restrictive word, leading to incorrect answers.

Across the four representative questions examined, a consistent picture emerges: incorrect answers were not random. They clustered around specific types of difficulty, modal qualification, normative recency, compound distractors, and ambiguous item formulation that cut across models and prompting strategies. This suggests that the errors reflect structural properties of both the models and the questions, rather than isolated stochastic failures. The analysis of these four cases helps answer RQ3.1 and RQ3.2 by showing that incorrect responses were not random hallucination events, but recurring failure patterns that can be categorized.

For RQ3.1, the selected cases suggest model-side limitations related to overgeneralization, difficulty handling modal verbs, and difficulty evaluating compound alternatives in multi-select questions.

For RQ3.2, the selected cases show that question formulation can also contribute to incorrect answers. Questions that ignore internal contradictions of the official reference material (such as the rule clash between Product Backlog and Sprint Backlog) capture the model's algorithmic attention to the wrong rule. Additionally, the tricky wording typical of certifications—which uses subtle restrictive adverbs to invalidate popular premises—exposes the fragility of LLMs when faced with assessments that prioritize lexical rigor over the inference of generalist conceptual knowledge. Together, these dimensions help explain why high aggregate accuracy can coexist with systematic errors in specific Scrum certification-style questions.

\FloatBarrier
\section{Implications}
\label{sec:implications}

This study has implications for evaluating LLMs in Software Engineering, for Scrum-style certification assessments, and for future research on domain-specific question-answering tasks.

\textbf{Implications for LLM Evaluation in Software Engineering}
The results show that aggregate accuracy alone is insufficient to characterize LLM behavior in normative Software Engineering domains. Although all evaluated models achieved high overall performance, the analyses revealed meaningful differences across models, prompting the development of strategies, Scrum topics, and question formats. This suggests that evaluations of LLMs in Software Engineering should go beyond reporting a single performance score. Stability across repeated executions, sensitivity to task characteristics, and recurring error patterns provide additional evidence about when model outputs are more or less dependable.

The findings also indicate that model choice had a stronger effect on performance than prompt choice. Structured prompts produced modest gains in some conditions, especially when they encouraged reasoning or grounding in the \emph{Scrum Guide}, but they did not eliminate differences among models. This reinforces the need to evaluate models under comparable prompting conditions rather than assuming that prompt engineering alone can compensate for differences in model capability.

\textbf{Implications for Scrum Certification-Style Assessment.}
Scrum certification-style questions provide a useful setting for evaluating LLMs because they combine objective scoring with normative domain knowledge. In this study, the models performed better on topics grounded in explicit terminology and well-delimited concepts, such as Artifacts, Empiricism, and Product Value. In contrast, weaker or more variable performance was observed in topics such as Scrum Values, Self-Managing Teams, and Stakeholders \& Customers, where questions often require finer distinctions among responsibilities, values, and stakeholder relationships.

Question format also shaped model performance. Single-answer multiple-choice questions were the most favorable format, whereas multi-select and True/False questions introduced additional difficulty. Multi-select items require the model to identify a complete set of correct alternatives while avoiding plausible distractors. True/False items can be affected by restrictive terms or subtle normative distinctions that invert the expected answer. These findings suggest that certification-style benchmarks should report performance by question format rather than treating all items as equivalent.

\textbf{Implications for Question and Benchmark Design.}
The qualitative analysis showed that incorrect answers were not random. Some errors were associated with model-side limitations, such as overgeneralization, difficulty handling modal verbs, and difficulty evaluating compound alternatives. Other errors were associated with ambiguity, underspecification, or competing interpretations in the question formulation. This distinction is important because an incorrect model answer may reveal not only a model limitation, but also a weakness in the assessment item itself.

For researchers designing benchmarks, these results suggest that question banks should be inspected not only for coverage and answer-key correctness, but also for ambiguity, restrictive wording, and dependence on narrow interpretations of reference material. In normative domains such as Scrum, small linguistic markers can substantially change the expected answer. As a result, qualitative error analysis can complement quantitative scoring by identifying whether failures arise from model behavior, question design, or their interaction.

\textbf{Implications for Future Research.}
The study design can be expanded to encompass other Agile frameworks, such as SAFe, LeSS, Nexus, and Kanban, with the aim of verifying whether the patterns identified in Scrum persist in more diverse and complex contexts. Future work should also incorporate additional models, such as Claude, enabling comparative analyses across different architectures and training strategies. Another promising direction involves the use of metamorphic testing to evaluate the robustness of models under controlled reformulations of questions, helping to distinguish between genuine conceptual understanding and superficial sensitivity to textual phrasing. Finally, it is recommended to expand the dataset to include other levels and types of certification, such as PSM II and PSPO, allowing for a more comprehensive evaluation of performance in scenarios that require greater depth of knowledge and contextual judgment.

Scrum certification-style questions constitute a controlled empirical setting for studying LLM behavior in normative Software Engineering tasks. Such settings enable analysis not only of whether models answer correctly, but also of how stable their answers are, where performance degrades, and which types of errors persist despite high aggregate accuracy.

\FloatBarrier
\section{Threats to Validity}
\label{sec:threats}

Threats to validity are discussed following the classification proposed by Wohlin et al.~\cite{wohlin2012experimentation}.

\textbf{Conclusion validity.}
The quantitative analyses rely on exact-match accuracy computed over balanced subsamples, which limits the precision of topic and format-level comparisons. The balanced samples of 104 and 96 questions were designed to reduce category imbalance. Still, the small number of items per category (eight per topic and 32 per format) means that individual question characteristics can disproportionately influence observed results. The qualitative error analysis necessarily involves interpretive judgment: although it was conducted collaboratively by the authors using traceable question identifiers and explicit analytic categories, the selection of representative cases and the assignment of errors to categories (model-side vs.\ question-formulation) involves subjectivity that cannot be fully eliminated. The interpretations were cross-checked with the dataset provider for cases involving potential gold-standard inconsistencies, thereby reducing but not eliminating this threat.

\textbf{Construct validity.}
Model performance was operationalized primarily as answer correctness relative to predefined gold-standard responses. This operationalization is appropriate for certification-style question answering. Still, it does not capture all aspects of model behavior, such as explanation quality, partial correctness, or the plausibility of alternative interpretations. This limitation is especially relevant for multi-select items and for questions involving subtle distinctions among Scrum accountabilities, values, and stakeholder relationships. To mitigate this threat, we complemented the quantitative analysis with a qualitative examination of error patterns, which helped distinguish more objective failures from cases involving ambiguity, underspecification, or plausible alternative interpretations. In addition, the dataset was derived from a PSM I preparation question bank and reviewed by Scrum-certified experts, which increases its alignment with the intended assessment context.

\textbf{Internal validity.}
All questions were executed independently, with no conversational memory across runs, to reduce cross-question contamination. Because LLM outputs may vary stochastically, each model--prompt combination was executed five times over the full dataset. The low variability observed across repeated runs reduces the likelihood that the main findings are artifacts of isolated executions. Nevertheless, some threats remain. First, the observed results may have been influenced by hidden model-side changes, service-side variability, or undocumented differences in inference behavior across providers. Second, the correctness of the gold-standard answers remains a potential source of error, particularly in questions that admit multiple plausible interpretations or depend on narrow certification-oriented readings. Third, the qualitative error analysis necessarily involves interpretive judgment, even though it was conducted using traceable question identifiers, explicit analytic categories, and discussion among the authors.

\textbf{External validity.}
The dataset consists of 993 English Scrum certification-style questions aligned with the \emph{Scrum Guide} (2020), which limits generalization to other Agile frameworks, certification ecosystems, languages, or future revisions of Scrum. Likewise, the evaluated models (\textit{GPT-5 mini}, \textit{Gemini 3 Flash}, and \textit{DeepSeek Chat 3.2}) represent a relevant but limited sample of contemporary LLMs. The dataset is also exclusively in English, which limits generalization to certification-style assessments conducted in other languages, including Portuguese, which is directly relevant to the Brazilian Software Engineering community. LLM performance in normative domains may vary across languages due to differences in training data distribution and the availability of Scrum-related reference material in non-English corpora. Results may differ for other models, providers, prompting configurations, inference settings, or interaction modes. Nevertheless, the study design is replicable and may be adapted to other Software Engineering domains that combine canonical knowledge, professional standards, and certification-style evaluation.

\section{Final Remarks}
\label{sec:final}

This paper evaluated three contemporary LLMs, \textit{GPT-5 mini}, \textit{Gemini 3 Flash}, and \textit{DeepSeek Chat 3.2}, on Scrum certification-style questions aligned with the Professional Scrum Master I (PSM I) assessment format. Using 993 questions based on the \emph{Scrum Guide} (2020), we analyzed model accuracy under different prompting strategies, examined intra-model variability across repeated executions, and investigated performance differences by Scrum topic, question format, and recurring error pattern.

Gemini 3 Flash achieved the strongest results across prompting strategies, while GPT-5 mini and DeepSeek Chat 3.2 also exceeded the PSM~I passing threshold under all conditions. Intra-model variability was low, indicating stable behavior across repeated executions. However, performance was not uniform. The models performed better on normatively explicit topics and single-answer multiple-choice questions, while showing greater fragility on topics such as Scrum Values, Self-Managing Teams, and Stakeholders \& Customers, as well as on multi-select and True/False items. Structured prompting produced modest gains in some conditions, but model choice had a stronger effect on accuracy than prompt choice.

These findings carry implications at three levels. For LLM evaluation in Software Engineering, they reinforce that aggregate accuracy should be complemented by analyses of stability, task sensitivity, and recurring failure modes. For practitioners and educators using Scrum certification-style assessments, they suggest that model choice matters more than prompt choice. That performance on normatively explicit topics should not be taken as evidence of reliability across all Scrum areas. For benchmark and question designers, the qualitative results indicate that some incorrect answers reflect weaknesses in question formulation rather than model limitations, motivating a systematic review of items that involve restrictive wording, competing interpretations, or tensions between general rules and event-specific exceptions in the \emph{Scrum Guide}.

The qualitative analysis further showed that incorrect answers followed recurring patterns rather than random failures. Some errors were associated with model-side limitations, such as overgeneralization, difficulty handling restrictive wording, and difficulty evaluating compound alternatives. Other errors were associated with ambiguity, underspecification, or competing interpretations in the questions themselves. Overall, these findings reinforce that high aggregate accuracy is not sufficient to characterize LLM behavior in certification-style Software Engineering tasks. 

The study has limitations that should be considered when interpreting these findings. The dataset consists of 993 English questions from a single PSM I preparation question bank, which limits generalization to other languages, certification ecosystems, or future revisions of the \emph{Scrum Guide}. The balanced subsamples used for topic- and format-level analyses contain only a small number of items per category, potentially leading to disproportionate influence from individual question characteristics on the observed results. The qualitative error analysis, although conducted collaboratively and cross-checked with the dataset provider, involves interpretive judgment that cannot be fully eliminated.

Future work should extend this design to other Agile frameworks, certification ecosystems, and normative Software Engineering domains, and should evaluate additional models, prompting strategies, and grounding mechanisms.

\FloatBarrier
\section*{Artifacts Availability}
\label{sec-artifact}

The primary database, the prompts used in this study, and the resulting datasets containing the Large Language Model (LLM) responses are publicly available at \url{https://doi.org/10.6084/m9.figshare.32087001}. 

\FloatBarrier
\section*{Acknowledgements}

The authors used ChatGPT to support language editing, clarity, readability, and minor textual refinement during manuscript preparation. No AI tool was used to generate scientific content, define the study design, select or screen studies, extract data, perform data analysis, interpret results, or draw conclusions. All AI-assisted text was reviewed, revised, and validated by the authors, who take full responsibility for the content of this manuscript.

This work has been partially funded by the project ``\textit{iSOP Base: Investigação e desenvolvimento de base arquitetural e tecnológica da Intelligent Sensing Operating Platform (iSOP)}'' supported by CENTRO DE COMPETÊNCIA EMBRAPII VIRTUS EM HARDWARE INTELIGENTE PARA INDÚSTRIA - VIRTUS-CC, with financial resources from the PPI HardwareBR of the MCTI grant number 055/2023, signed with EMBRAPII.

\bibliographystyle{ACM-Reference-Format}
\bibliography{samples/sample-base}

\end{document}